\documentclass[pra,aps,longbibliography,twocolumn,superscriptaddress,floatfix]{revtex4-1}
\usepackage{amsmath,amssymb,graphicx,units}
\usepackage[usenames,dvipsnames]{color}
\usepackage{verbatim}
\usepackage{lipsum}
\usepackage{setspace}
\usepackage{dcolumn}
\usepackage{bm}
\usepackage[bookmarks=false,linkcolor=blue,urlcolor=blue,colorlinks,citecolor=blue]{hyperref}

\usepackage{mathrsfs}
\usepackage[T1]{fontenc}

\begin{document}

\title{Superior resilience to poisoning and amenability to unlearning in quantum machine learning}

\author{Yu-Qin Chen}
\email{yqchen@gscaep.ac.cn}
\affiliation{Graduate School of China Academy of Engineering Physics, Beijing 100193, China}

\author{Shi-Xin Zhang}
\email{shixinzhang@iphy.ac.cn}
\affiliation{Institute of Physics, Chinese Academy of Sciences, Beijing 100190, China}

\date{\today}

\begin{abstract}

The reliability of artificial intelligence hinges on the integrity of its training data, a foundation often compromised by noise and corruption. Here, through a comparative study of classical and quantum neural networks on both classical and quantum data, we reveal a fundamental difference in their response to data corruption. We find that classical models exhibit brittle memorization, leading to a failure in generalization. In contrast, quantum models demonstrate remarkable resilience, which is underscored by a phase transition-like response to increasing label noise, revealing a critical point beyond which the model's performance changes qualitatively.
We further establish and investigate the field of quantum machine unlearning, the process of efficiently forcing a trained model to forget corrupting influences. We show that the brittle nature of the classical model forms rigid, stubborn memories of erroneous data, making efficient unlearning challenging, while the quantum model is significantly more amenable to efficient forgetting with approximate unlearning methods. Our findings establish that quantum machine learning can possess a dual advantage of intrinsic resilience and efficient adaptability, providing a promising paradigm for the trustworthy and robust artificial intelligence of the future.
\end{abstract}

\maketitle


\section*{Introduction}

The remarkable success of artificial intelligence (AI) has placed it as a transformative force in science and society~\cite{lecun2015deep, jordan2015machine, Vaswani2017attention}, yet the reliability of these powerful models is fundamentally contingent on the integrity of their training data. In real-world applications, datasets are often compromised by corruption, such as mislabeled examples or malicious poisoning attacks, which can severely impair a model's generalization, introduce dangerous biases, and create significant security vulnerabilities~\cite{barreno2006can, goldblum2022dataset,Tian2023survey, Zhao2025survey}. A closely related and increasingly urgent challenge is that of machine unlearning: the need to efficiently remove the influence of specific data from a trained model to comply with privacy regulations or to correct for erroneous information~\cite{Cao2015mu, Ginart2019datadelete, Bourtoule2021, Wang2024musurvey, Goel2024corrective, Liu2025llmunlearning, Qu2025surveyllm}. While retraining a model from scratch on a sanitized dataset offers a definitive solution, its prohibitive computational cost for large-scale systems makes it impractical, driving the search for more efficient alternatives.

As the field of quantum computing evolves rapidly~\cite{Preskill2018nisq, Bharti2021review, Cerezo2020breview}, quantum machine learning (QML) has emerged as a promising new paradigm with the potential to solve problems intractable for classical computers~\cite{Biamonte2017a, Lloyd2013, Cai2015prl, Havlicek2019nature, Gao2018sa, Li2017control, Huang2020powerdata, Huang2022science, Zhang2021vqnhe, Chen2022alpha,Cerezo2022ncs, Wang2024review}. These models, often implemented as quantum neural networks (QNNs), leverage principles like superposition and entanglement to navigate vast computational spaces~\cite{Schuld2014quest, Romero2017qae, Liu2018born, Benedetti2019qst, Beer2020dnn, Shen2020infor, Dallaire-Demers2018qgan, Cong2019, Prez-Salinas2020reupload, Banchi2021qi,Zhang2021nqas, Li2022recent, Zheng2023fqnn, Jger2023nc, Miao2023nnvqa, Zhang2024diffusion, Sein2025_invariant, Zhang2025qec}. However, despite rapid theoretical and experimental progress, the behavior of quantum models in the face of real-world data imperfections remains a critical and largely unexplored frontier. While some studies have investigated the vulnerability of QNNs to inference-time adversarial attacks \cite{Huang2011, Biggio2013, Kurakin2017, Lu2020,Liu2020,Du2021,Weber2021,Liao2021, Ren2022, Gong2021nsr, West2023a, West2023}, their response to training-time data corruption is poorly understood. This distinction is critical: adversarial examples challenge a model's perception at inference, whereas data poisoning corrupts the model's fundamental knowledge during training. Furthermore, the subsequent problem of how to correct a poisoned model, a field we establish here as quantum machine unlearning, remains an entirely unexplored area. This gap presents a formidable barrier to developing quantum AI that is not only powerful but also trustworthy and secure.

\begin{figure*}[t]\centering
	\includegraphics[width=\textwidth]{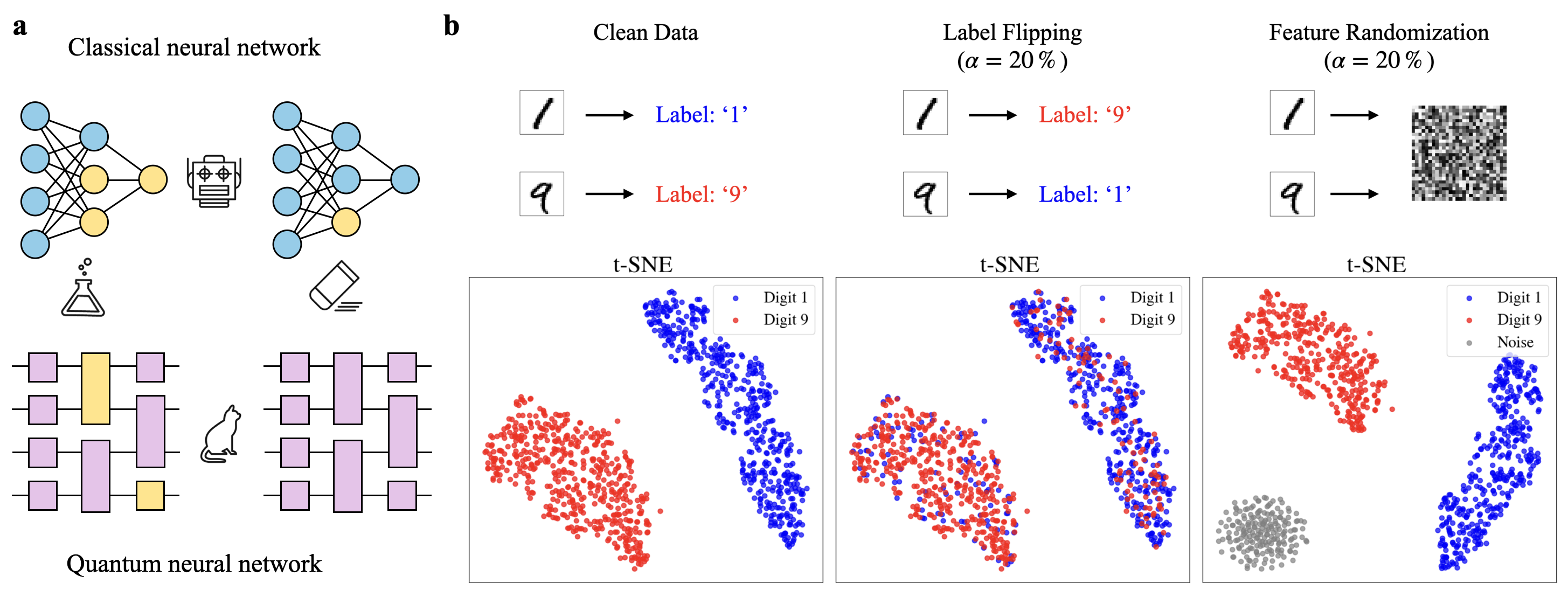}
	\caption{\textbf{Conceptual Framework and Data Corruption Protocols.}
    \textbf{a}, Conceptual illustration of the learning and unlearning lifecycle investigated in this work. Both the classical neural network and the quantum neural network are subjected to training-time data corruption (poisoning) and subsequent correction (machine unlearning). The distinct icons for each model represent their different computational paradigms. QNN shows superior performance for learning under data corruption and unlearning.
    \textbf{b}, Visualization of the data corruption protocols on the MNIST dataset, illustrated with representative examples and t-SNE low-dimension projections of the feature space. The leftmost column shows the clean data, where the t-SNE plot reveals two well-separated clusters for digits `1' (blue) and `9' (red). The middle column illustrates Label Flipping with a noise ratio of $\alpha=0.2$, creating in-cluster anomalies as points retain their geometric position but adopt the opposing class's label. The rightmost column shows Feature Randomization ($\alpha=0.2$), which replaces input images with random vectors, introducing a new, structurally distinct cluster of noise points (grey).}
\label{fig:setup}
\end{figure*}

In this work, we bridge these critical gaps through a systematic, comparative study of a classical machine learning model of multi-layer perceptron (MLP) and a QNN, evaluating their response to different data corruption and different unlearning approaches across both classical and quantum datasets (Fig.~\ref{fig:setup}). We reveal a fundamental difference in their learning mechanisms. The classical model exhibits brittle memorization, attempting to fit all data points, which leads to a catastrophic collapse in generalization when faced with contradictory information. In stark contrast, the QNN demonstrates a superior intrinsic resilience, prioritizing the general data structure over statistical outliers. This robust behavior is underscored by a distinct phase transition-like response to increasing label noise. This suggests that the QNN's learning dynamics is not a simple fitting exercise but a complex system exhibiting critical phenomena, a behavior far more structured than the continuous performance degradation of the MLP.
Building on this discovery, we investigate the field of \textbf{quantum machine unlearning}. We find that the brittle nature of the classical model forms deep, stubborn memories that make efficient unlearning profoundly challenging. Conversely, the quantum model displays remarkable model plasticity, proving highly amenable to efficient forgetting. Approximate unlearning methods are not only more stable but can achieve superior performance to a retrain from scratch within the same training time window. This work systematically demonstrates the dual advantage for QML by providing both a foundational analysis of controlled levels of data corruption in QML and the first-ever framework for quantum machine unlearning, thereby paving the way for the development of secure and adaptable quantum technologies.

\section*{Results}

\begin{figure*}[t]\centering
	\includegraphics[width=\textwidth]{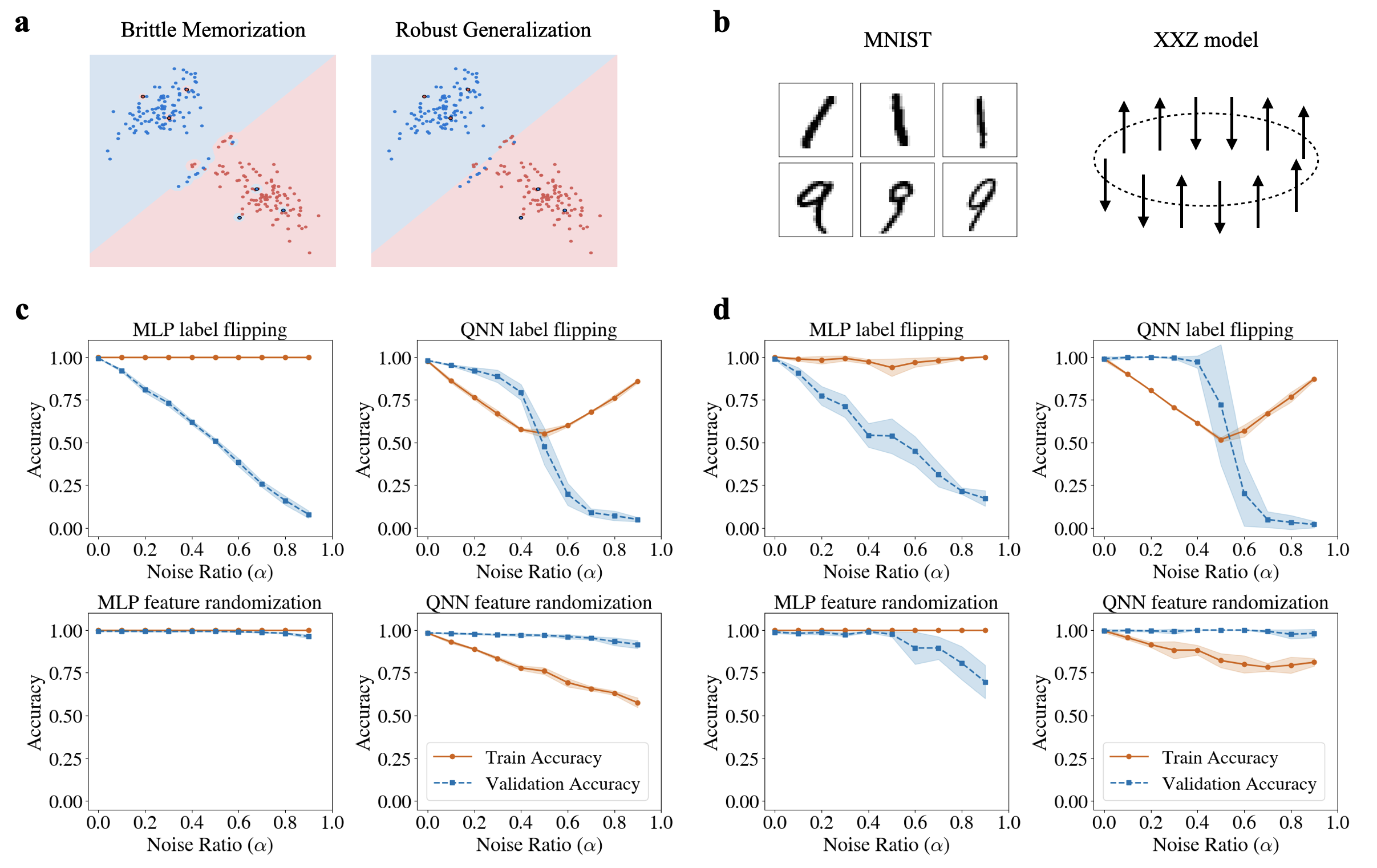}
	\caption{\textbf{Superior Resilience of Quantum Models to Data Corruption.}
    \textbf{a}, Conceptual illustration of the two learning paradigms. The classical MLP engages in brittle memorization, deforming its decision boundary to fit mislabeled outliers (e.g., blue dots in red region). The QNN exhibits robust generalization, maintaining a simple boundary and correctly classifying the majority of points, even at the cost of misclassifying the wrong outliers.
    \textbf{b}, Representative data from the two classification tasks. Left: distinguishing MNIST digits `1' and `9'. Right: distinguishing ground state phases of the 1D XXZ spin model.
    \textbf{c}, Empirical results for the classical MNIST dataset. Plots show training (orange solid) and validation (blue dashed) accuracy versus the data noise ratio $\alpha$. Under label flipping, the MLP's validation accuracy continuously degrades, while the QNN maintains a robust performance plateau before a sharp critical transition. Both models are robust to feature randomization with moderate $\alpha$. Shaded regions represent the standard deviation over 5 optimization runs.
    \textbf{d}, Empirical results for the quantum XXZ dataset. The trends are consistent with the quantum data, showing the MLP's continuous compromise versus the QNN's robust plateau under label flipping. The QNN's resilience is thus a general feature, independent of the nature of the data.}
\label{fig:learn}
\end{figure*}

To systematically investigate the behavior of machine learning models in corrupted environments, we conducted a comparative study centered on a classical MLP and a QNN built on top of parameterized quantum circuits (PQC). These models were evaluated on two distinct binary classification tasks (Fig.~\ref{fig:learn}\textbf{b}): a classical task using MNIST handwritten digits (`1' vs `9') and a quantum task classifying the ground state phases of the one-dimensional XXZ spin Hamiltonian. We first probe the intrinsic resilience of each model by corrupting a controlled fraction $\alpha$ of the training data using two distinct protocols: \textbf{Label Flipping} and \textbf{Feature Randomization} (Fig.~\ref{fig:setup}\textbf{b}). The former introduces direct contradictions by inverting a sample's class label, while the latter destroys the data's inherent structure by replacing the entire feature vector with random noise.  Subsequently, we assess each model's capacity for repair by applying four different machine unlearning algorithms to remove the influence of the corrupted data. The following results reveal a consistent dual advantage for the quantum model, demonstrating superior resilience to training data corruption and greater amenability to efficient forgetting.

\subsection{Superior Resilience of Quantum Models to Graded Data Corruption}

\subsubsection*{Label Flipping: Robust Plateau versus Continuous Compromise}
In the Label Flipping scenario, where a fraction $\alpha$ of training labels are inverted, the MLP and QNN exhibit starkly different resilience strategies (Fig.~\ref{fig:learn}\textbf{c}, \textbf{d} (top)). These strategies reveal fundamental differences in their ability to distinguish signal from noise.

The classical MLP demonstrates a high sensitivity to label noise, resulting in a continuous compromise of its generalization ability. As shown in Fig.~\ref{fig:learn}\textbf{c}, \textbf{d} (top left), its validation accuracy begins to degrade almost immediately and steadily with any increase in the noise ratio $\alpha$. This behavior stems from its tendency towards brittle memorization as illustrated in Fig.~\ref{fig:learn}\textbf{a}. MLP attempts to accommodate every contradictory data point, causing a steady erosion of the global decision boundary's integrity. While its training accuracy remains high, this comes at the direct cost of its performance on unseen data, indicating an inability to effectively disregard even low levels of noise.

In stark contrast, the QNN displays a far superior strategy characterized by a robust performance plateau followed by a sharp, critical transition at $\alpha=0.5$. As shown in Fig.~\ref{fig:learn}\textbf{c},\textbf{d} (top right), the system exhibits two distinct phases: a signal-dominated phase for $\alpha < 0.5$, where validation accuracy remains remarkably high, and a noise-dominated phase for $\alpha > 0.5$, where performance collapses. Within the signal-dominated phase, the QNN effectively ignores the noisy outliers—a behavior confirmed by its decreasing training accuracy—thereby preserving the integrity of its learned representation. The sharp drop at the critical point $\alpha_c \approx 0.5$ is characteristic of a phase transition. This behavior is the hallmark of a robust system that maintains its ordered state (correct classification) against disorder (label noise) up to a definitive transition point. In other words, the label noise is a relevant perturbation for classical MLP but an irrelevant perturbation for QNN. 
This advantage is not merely about model size, as it persists even when comparing QNNs of varying depth against MLPs of varying capacity (see Supplemental Materials).
The ability of QNN to maintain a predictable window of high performance under a significant degree of contamination makes it an inherently more reliable model for real-world applications where training data is inevitably imperfect.

\subsubsection*{Feature Randomization: Robustness to Out-of-Distribution Noise}
In the Feature Randomization scenario, where input feature vectors are replaced with random noise, both models exhibit considerable robustness, as the out-of-distribution nature of the corrupted data is generally easy to distinguish. However, a closer examination, particularly in data-limited regimes, reveals a more nuanced picture and a distinct advantage for the QNN.

On the smaller XXZ dataset, the MLP's performance degrades notably at high corruption ratios ($\alpha > 0.8$), where the number of remaining clean samples becomes low (Fig.~\ref{fig:learn}\textbf{d} (bottom left)). The model struggles to form a stable decision boundary amidst a sea of random noise. In stark contrast, the QNN's validation accuracy remains almost entirely unaffected across all values of $\alpha$ (Fig.~\ref{fig:learn}\textbf{d} (bottom right)). This indicates a superior generalization ability for QML models~\cite{caro2022generalization} to identify and completely disregard out-of-distribution noise, even when clean training examples are extremely scarce.

This performance difference is less apparent on the larger MNIST dataset (Fig.~\ref{fig:learn}\textbf{c} (bottom)), as even a high corruption ratio leaves a sufficient absolute number of clean samples for the MLP to learn the task effectively. Therefore, while both architectures can handle this type of noise, the QNN's performance under Feature Randomization, especially in the data-scarce limit, provides further evidence of its intrinsic robustness.

\begin{figure*}[t]\centering
	\includegraphics[width=\textwidth]{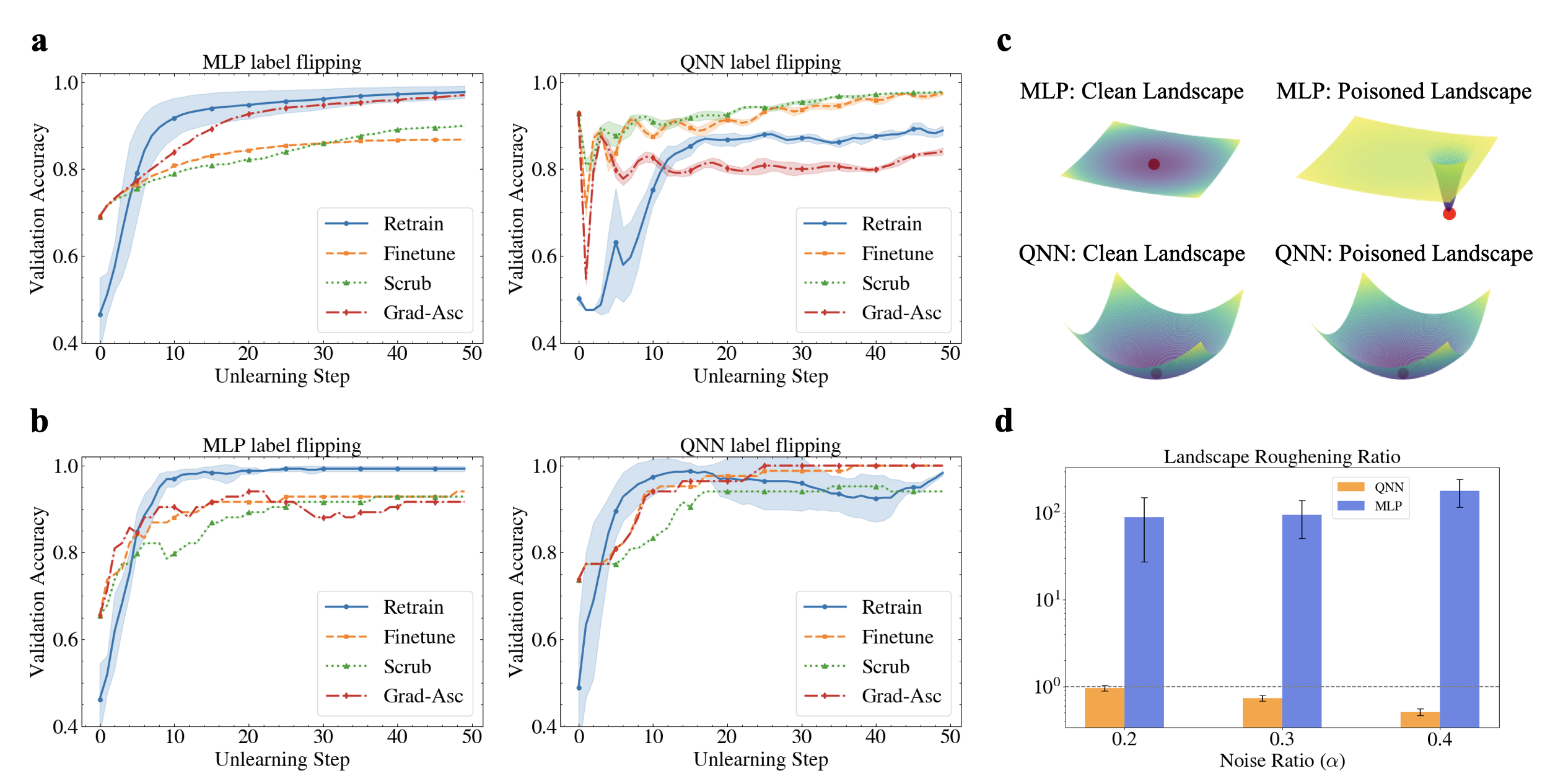}
	\caption{\textbf{Superior Unlearning Plasticity of Quantum Models and its Geometric Origin.}
    \textbf{a, b}, Unlearning dynamics for the MLP and QNN on the classical MNIST dataset (\textbf{a}) and the quantum XXZ dataset (\textbf{b}), following training on data corrupted by label flipping ($\alpha=0.3$). The plots show validation accuracy versus the unlearning step for four different unlearning methods. For the MLP, the Retrain method serves as a clear upper bound, while computationally cheaper methods consistently underperform, evidencing stubborn memories. For the QNN, approximate unlearning methods are highly effective, often outperforming the costly Retrain baseline within the early training time window after a transient performance dip, demonstrating remarkable model plasticity. Shaded regions represent the standard deviation over 5 runs.
    \textbf{c}, Schematic illustration of the loss landscape geometry. The MLP's landscape deforms from a wide, flat minimum (Clean) into a pathologically sharp, brittle minimum (Poisoned) to memorize outliers. The QNN's robust, stable minimum remains largely unperturbed by the data corruption.
    \textbf{d}, Quantitative analysis via the LRR, defined as the ratio of the Hessian trace in the noisy vs. clean scenarios. The MLP's LRR is several orders of magnitude greater than one, confirming its landscape's fragility. The QNN's LRR is consistently near unity (dashed line), providing direct evidence of its landscape's structural stability. Error bars represent the standard deviation of the LRR across different initializations.}
\label{fig:unlearn}
\end{figure*}

\subsection{Quantum Machine Unlearning: Plasticity versus Stubborn Memories}
Building on the finding of the QNN's superior resilience, we now assess its capacity for repair by establishing the first systematic investigation into quantum machine unlearning. Our unlearning protocol begins with a model initially trained on a dataset $D_{\text{train}}$, which is a union of a clean \textbf{retain set}, $D_{\text{retain}}$, and a polluted \textbf{forget set}, $D_{\text{forget}}^{\text{polluted}}$. The goal of unlearning is to efficiently alter the trained model to approximate a new model trained from scratch solely on $D_{\text{retain}}$ with sufficient training steps, thus erasing the influence of the corrupted data. A separate, held-out clean \textbf{validation set}, $D_{\text{val}}$, is used to evaluate the final performance of the unlearned model.

To this end, we compare three efficient, approximate unlearning strategies as well as the retrain from scratch setting:
\begin{itemize}
    \item \textbf{Retrain:} The baseline method that discards the polluted model entirely and trains a new one from scratch using only $D_{\text{retain}}$, also known as exact unlearning. It is effective but prohibitively costly in practice.
    \item \textbf{Finetune:} The simplest approximate method, which takes the polluted model and continues training on $D_{\text{retain}}$ with the goal of overwriting the bad knowledge as implied by catastrophic forgetting \cite{French1999cf, Goel2023toward}.
    \item \textbf{Scrub:} A multi-objective technique involving a ``teacher'' (the original polluted model) and a ``student'' (the new model being unlearned). It simultaneously trains on $D_{\text{retain}}$ while encouraging the student's outputs on the forget set to diverge from those of the teacher and outputs on the retain set to be consistent with those of the teacher, thus actively forgetting \cite{Kurmanji2023scrub}.
    \item \textbf{Gradient Ascent:} A reverse learning approach that modifies the optimization objective to perform standard gradient descent on the retain set loss while simultaneously performing gradient ascent on the forget set loss, actively unlearning the erroneous samples by reversing the training process \cite{Trippa2024ga}.
\end{itemize}
Our unlearning studies focus on the Label Flipping scenario, as it presents the most relevant and challenging test case. In the Feature Randomization scenario, the model's performance is not significantly compromised at the low-to-moderate corruption ratios ($\alpha$) typical for unlearning studies. The model effectively learns to ignore these out-of-distribution outliers, meaning the prerequisite for unlearning—a genuinely corrupted model in need of repair—is not met. Conversely, studying unlearning at very high corruption ratios ($\alpha > 0.5$) would shift the problem from ``unlearning a small data subset'' to ``few-shot relearning from a tiny retain set'', a fundamentally different paradigm. Therefore, Label Flipping at moderate $\alpha$ represents the most suitable testbed where the model is sufficiently damaged to necessitate repair.

The unlearning advantage of the QNN is not merely theoretical but profoundly practical. We observe a stark contrast in the unlearning dynamics between the classical and quantum models.
For the classical MLP, the Retrain method quickly establishes a high-performance baseline. However, the computationally cheaper approximate unlearning methods consistently underperform, struggling to close the performance gap (Fig.~\ref{fig:unlearn}\textbf{a}, \textbf{b} (left)). This suggests that the MLP forms ``stubborn memories'' of the poisoned data,  which cannot be easily erased without a costly hard reset via full retraining. 

Conversely, for the QNN, the Retrain process from a random initialization exhibits slower convergence and greater instability. Strikingly, the approximate unlearning methods are not only more stable but can achieve superior validation accuracy within the same limited computational window than retrain from scratch (Fig.~\ref{fig:unlearn}\textbf{a, b} (right)). This demonstrates the QNN's remarkable ``unlearning plasticity''. The pre-trained state, even though poisoned, serves as a robust starting point that is highly amenable to efficient forgeting. This highlights a more favorable trade-off between computational cost and performance recovery in the quantum model, marking a key practical advantage for quantum unlearning. 

An interesting feature of the QNN's unlearning dynamics, particularly on the MNIST dataset, is the initial, transient dip in validation performance before its subsequent recovery and ascent (Fig.~\ref{fig:unlearn}\textbf{a} (right)). We interpret this not as a flaw, but as evidence of a necessary unanchoring process. The initial poisoned state represents a stable but incorrect local minimum. The unlearning procedure acts as a targeted perturbation that temporarily increases the model's loss, effectively heating the system to allow for broader exploration of the loss landscape. This transient destabilization enables the QNN to escape the suboptimal minimum before it reconverges to a better, more correct solution. The MLP's unlearning curves lack this dynamical pattern, suggesting its stubborn memories correspond to a much deeper, inescapable minimum.

\subsection{Mechanism of Resilience: Loss Landscape Geometry}
To understand the underlying mechanism behind the QNN's superior resilience and unlearning plasticity, we investigated the geometry of the loss landscapes found by both classical and quantum models. The flatness of a minimum in the loss landscape, which is inversely related to its sharpness, is a key indicator of good generalization and robustness to perturbations \cite{Hochreiter1997, Keskar2017}. We characterized this geometry by computing the spectrum of the Hessian matrix ($\nabla^2 \mathcal{L}$) of the loss function at the final trained model parameters.

While the absolute values of the Hessian eigenvalues are difficult to compare directly between models of such different architectures and parameter counts, their relative change in response to data corruption provides a clear and normalized measure of landscape stability. To quantify this, we define the \textbf{Landscape Roughening Ratio} (LRR) as the ratio of the Hessian matrix trace of the model trained on polluted data with that of the model trained on clean data:
\begin{equation}
    \text{LRR} = \frac{\text{Tr}(\mathbf{H}_{\text{noisy}})}{\text{Tr}(\mathbf{H}_{\text{clean}})}
    \label{eq:lrr}
\end{equation}
An LRR close to 1 indicates a highly robust landscape that is insensitive to noise, whereas a large LRR signifies a fragile landscape that deforms significantly toward sharpness to memorize corrupted data in a brute-force way.

Our analysis reveals a stark difference in landscape stability (Fig.~\ref{fig:unlearn}\textbf{c, d}). The classical MLP exhibits extreme fragility, with an extremely large LRR ($O(10^2)$). This astronomical increase confirms that the MLP's landscape undergoes a violent transformation to memorize the contradictory noisy labels, shifting from a very flat basin to a pathologically sharp and brittle minimum. Escaping such a sharp valley poses challenges on the unlearning algorithm. Approximate methods might fail because their gentle gradient updates are insufficient to exit this pathological basin.

In stark contrast, the QNN demonstrates exceptional stability. Its Hessian trace remains similar, yielding an LRR of approximately or smaller than 1. The QNN's architecture and optimization dynamics appear to resist the formation of the sharp, complex minima required to overfit to outliers. It fundamentally preserves the clean landscape geometry, thereby maintaining its robust generalization performance. This inherent structural stability, as captured by its near-unity LRR, is the origin of both its resilience to poisoning and its amenability to efficient unlearning. A fascinating open question arises when considering the scaling of these models. While deeper circuits may offer greater expressivity, they also risk the onset of barren plateaus—regions of extreme flatness \cite{McClean2018bp, Wang2020nibp, Cerezo2021hessianbp, Zhang2023fldc, Zhang2024predict, Larocca2025bpreview}. Future research must therefore explore a critical trade-off: does the flatness inherent to barren plateaus represent an ultimate form of robustness, or does it constitute a trivial stability that comes at the cost of all learnable features?

This conclusion is further supported by an analytical derivation on minimal classical and quantum models (see details in Supplementary Materials). Using a single-neuron MLP and a single-qubit quantum classifier, we analytically compute their Hessians. The derivation reveals that the MLP's curvature is built from strictly positive components. In contrast, the QNN's Hessian contains an additional data-error-phase interaction term which can become negative for mislabeled outliers. This provides a theoretical mechanism for curvature-balancing, where the influence of data noise on the landscape geometry is actively dampened, explaining the spectral stability we observe empirically.

\section*{Discussion}

Our findings show intrinsic differences of two learning paradigms. In the face of corrupted data, the classical MLP acts as a diligent but brittle stenographer, meticulously recording every detail, including falsehoods. The QNN, in contrast, behaves more like a discerning editor, prioritizing the integrity of the main narrative over accommodating every outlier. Our analysis of the loss landscape geometry provides a compelling mechanism for this difference, revealing that the QNN possesses a dual advantage rooted in a single, fundamental property: the inherent structural stability of its landscape.

Our central finding that QNNs exhibit a remarkable resilience to data corruption can be understood as the manifestation of the strong generalization properties for QML models~\cite{caro2022generalization}. The theoretical guarantee of a small generalization error is not merely a passive property for clean data; our results show that it is an active one. The QNN fights to preserve its generalizable solution even when pushed towards memorization by label noise. Our Hessian analysis reveals the physical mechanism that underpins this robust generalization. While the concept of flat minima is well-established for generalization, we introduce the crucial dimension of landscape stability under data corruption. The classical MLP finds an exceedingly flat but fragile minimum on clean data, which deforms into a pathologically sharp landscape under data noise. In contrast, the QNN's loss landscape is structurally stable, with its geometry remaining almost entirely unperturbed by data corruption, as quantified by its near-unity LRR. This stability is the direct cause of the favorable generalization guarantees for QML. The macroscopic phase transition we observe at $\alpha=0.5$ is the ultimate signal of this stability: the system resists disorder up to a critical point before undergoing a cooperative shift, a behavior far more structured and robust than the continuous performance compromise of the MLP.

Our finding that QNNs exhibit a robust resilience to label noise, preferring to maintain a simple, generalizable solution rather than memorizing outliers, may seem to be in tension with recent work demonstrating that QNNs possess a high capacity for memorization~\cite{Gil-Fuster2024generalization}. where QNNs is shown to be expressive enough to fit pure noise.
In fact, our work provides crucial complementary insight into the model's inductive bias. While Ref.~\cite{Gil-Fuster2024generalization} establishes what QNNs can do (capacity), our comparative study with MLPs reveals what they prefer to do. Our results show that when a dataset contains a mixture of signal and noise, QNN creates a strong preference for the signal. The MLP, on the other hand, can also remember all training samples~\cite{zhang2021understanding} while lacking this robust inductive bias and readily engages in brittle memorization of the noise.
Therefore, the QML generalization is twofold: QNNs possess a surprisingly high capacity to memorize, yet they simultaneously exhibit a built-in resistance to doing so when a simpler, more general solution exists. The high potential capacity coupled with a strong regularization bias is precisely what makes QNN promising for learning in noisy, real-world environments.

These results compel us to reconsider the nature of the quantum advantage in the near term. While the quest for exponential speedups remains an ultimate goal, our work suggests that a more immediate and practical advantage may lie in the domains of robustness and trustworthiness. In an information ecosystem rife with noise, bias, and adversarial manipulations, the ability of a model to resist corruption and be efficiently corrected is of great importance. The dual advantage of resilience and plasticity positions QML as a uniquely promising paradigm for building reliable AI systems.

However, our study carves out only the first piece of this vast territory. We operated under the assumption of fully known forget sets and noise at the data level. This opens several crucial future directions. A primary question is to disentangle the effects of data-level noise from device-level quantum noise. How does the inherent noise of near-term hardware interact with our observed resilience? Does quantum noise act as a further regularizer, enhancing robustness, or does it degrade the model's ability to distinguish signal from noise, potentially erasing this advantage?
Furthermore, future work should also explore these phenomena across a wider range of model architectures, datasets, learning tasks, and unlearning techniques. By continuing to chart the response of quantum models to real-world data imperfections, we can pave the way for new secure and adaptable quantum technologies.

\section*{Methods}

\subsection*{Classification Tasks}

In this study, we investigate two distinct binary classification problems to evaluate the performance of our methods.

{\textbf{XXZ Model Ground State Classification.}}
This problem originates from quantum many-body physics and involves classifying the quantum ground states of the one-dimensional spin-$1/2$ Heisenberg XXZ model. The system consists of a chain of $L=12$ spins with periodic boundary conditions. The Hamiltonian of the system is given by:
\begin{equation}
    H_{\text{XXZ}} = \sum_{i=1}^{L} \left( \sigma_i^x \sigma_{i+1}^x + \sigma_i^y \sigma_{i+1}^y + \Delta \sigma_i^z \sigma_{i+1}^z \right)
\end{equation}
where $\sigma_i^{\alpha}$ are the Pauli matrices for the spin at site $i$, and $\Delta$ is the anisotropy parameter that controls the phase of matter. The boundary condition imposes that $\sigma_{L+1} \equiv \sigma_1$.

The dataset is generated by numerically calculating the ground state eigenvector of this Hamiltonian for different values of $\Delta$ using exact diagonalization. The ground state, a normalized vector in $\mathbb{C}^{2^L}$, serves as the input feature vector $x$. The task is a binary classification problem to distinguish between two quantum phases based on the anisotropy $\Delta$:

\begin{itemize}
    \item \textbf{Class 0 (Gapless Phase)}: The ground states generated for values of the anisotropy parameter $\Delta$ in the range:
    \[ \Delta \in [-0.96, -0.94, \dots, 0.94, 0.96] \]
    This corresponds to the gapless (critical) phase of the XXZ model. 

    \item \textbf{Class 1 (Gapped Phase)}: The ground states generated for values of the anisotropy parameter $\Delta$ in the range:
    \[ \Delta \in [1.02, 1.04, \dots, 2.98, 3.00] \]
    This corresponds to the gapped antiferromagnetic Ising-like phase.
\end{itemize}

The validation dataset includes the ground states from $\Delta \in [-0.97, -0.95, \dots, 2.99]$.

{\textbf {MNIST Digit Classification.}}
This is a classical computer vision task based on a subset of the well-known MNIST dataset of $28\times 28$ handwritten digits. 

The task is a binary classification problem designed to distinguish between two specific digits, `1' and `9'. To create a balanced dataset, we select 250 image samples for each of these two digits. The input feature vector $x$ is the flattened pixel representation of the image. The classes are defined as follows:

\begin{itemize}
    \item \textbf{Class 0}: 250 images of the digit `1'.
    \item \textbf{Class 1}: 250 images of the digit `9'.
\end{itemize}
The validation dataset includes 1000 other images of the two digits.

\subsection*{Model architectures}
\label{sec:methods_models}
\begin{figure}[t]\centering
	\includegraphics[width=0.48\textwidth]{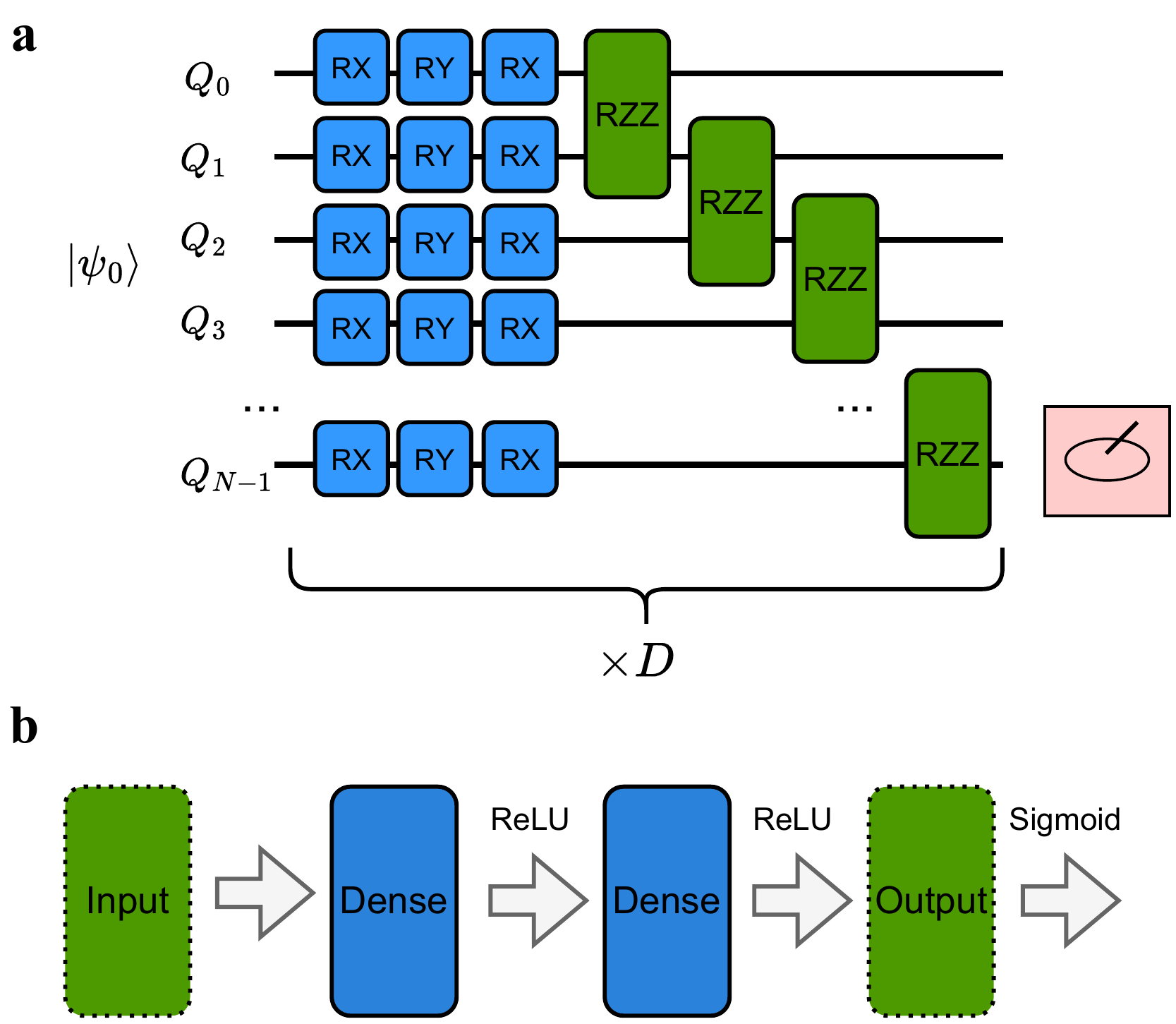}
	\caption{\textbf{Model Architectures.} 
    \textbf{a}, Schematic of the QNN architecture used in this study. The model consists of an N-qubit register initialized with an input state $|\psi_0\rangle$. A variational ansatz is applied, composed of $D$ repeated layers. Each layer consists of single-qubit rotations (RX, RY, RX) followed by entangling two-qubit gates (RZZ) applied to adjacent qubits. A final measurement is performed to extract the classical output. 
    \textbf{b}, Diagram of the MLP architecture. The classical model is a feed-forward neural network consisting of an input layer, two fully-connected (Dense) hidden layers each followed by a ReLU activation, and a final output layer with a sigmoid activation function for binary classification.}
\label{fig:architecture}
\end{figure}

We compare the classical MLP with QNN. The specific hyperparameters for each model were chosen to achieve effective learning on their respective tasks (see Supplemental Materials for details). The schematic architectures are shown in Fig.~\ref{fig:architecture}.

{\textbf{MLP.}} Our classical model is a standard feed-forward multi-layer perceptron, used for both the XXZ and MNIST tasks. The architecture consists of an input layer accepting the flattened data vectors, followed by two hidden layers. The Rectified Linear Unit (ReLU) serves as the activation function for both hidden layers. The output layer consists of a single neuron with a sigmoid activation function to produce a probability for the binary classification task. For the XXZ model ground state classification task, the wavefunction components are split into real and imaginary part for the pure real input of MLP.

{\textbf{QNN.}} Our quantum model is a parameterized quantum circuit, implemented using the {\sf TensorCircuit-NG} library \cite{Zhang2022tc}. The architecture of the PQC was adapted to the specific dataset to ensure effective learning.
\begin{itemize}
    \item For the XXZ task, the model operates on a register of 12 qubits, matching the spin chain length. The 4096-dimensional ground state wavefunction is loaded directly into the initial state via amplitude encoding. The variational ansatz has a depth of 4.
    \item For the MNIST task, the model operates on 10 qubits. The 1024-dimensional padded input vector is loaded via amplitude encoding. To handle the increased complexity of this classical dataset, the variational ansatz has a depth of 6.
\end{itemize}
For both tasks, each layer of the ansatz is composed of single-qubit rotations (RX, RY, and RX gates) applied to all qubits, followed by a layer of RZZ gates ($e^{i\frac{\theta}{2} Z_jZ_{j+1}}$) arranged in a linear chain to generate entanglement. The classification output is derived from the expectation value of the Pauli-Z operator measured on the final qubit, which is then passed through a scaled sigmoid function to map the output to a probability.

\subsection*{Data Contamination}

{\subsubsection*{Data Contamination Methods}}
To simulate real-world scenarios where training data may contain errors, we employ two distinct methods to contaminate an originally clean dataset.

\textbf{Label Flipping.}
This method targets the labels ($y$) of the data, simulating common annotation errors. For a binary classification task where the labels are $y \in \{0, 1\}$, the label flipping operation is defined as:
\begin{equation}
    y_{\text{poisoned}} = 1 - y_{\text{clean}}
\end{equation}
This directly corrupts the mapping between features and labels, rendering the model to learn an incorrect, and potentially opposite, decision boundary. It represents a challenging form of data poisoning.

{\textbf{Feature Randomization}}
This method targets the features ($x$) of the data, simulating corrupted or replaced input. For a subset of samples of ratio $\alpha$, the original feature vector $x \in \mathbb{C}^{d}$ is replaced by a randomly generated vector, $x'_{\text{poison}}$. The random vector is real for MNIST task and complex for XXZ model ground state task. For the complex one, the generation process is as follows:
\begin{enumerate}
    \item Two real-valued vectors, $\theta_{\text{real}}$ and $\theta_{\text{imag}}$, are drawn from a normal (Gaussian) distribution, where each component is sampled independently:
    \begin{equation}
        (\theta_{\text{real}})_j, (\theta_{\text{imag}})_j \sim \mathcal{N}(\mu=0, \sigma^2=1)
    \end{equation}
    \item The poisoned input vector is then constructed using these random vectors as its real and imaginary parts and finally get normalized:
    \begin{equation}
        x'_{\text{poison}} = \theta_{\text{real}} + i \cdot \theta_{\text{imag}}
    \end{equation}
\end{enumerate}
This method completely destroys the original feature structure, forcing the model to associate a label with meaningless random noise, which can severely damage the learned feature representations.

{\subsubsection*{Dataset Partition Definitions}}
In the context of our machine unlearning experiments, the data is partitioned into the following sets:

\begin{itemize}
    \item[\textbf{$\mathcal{D}_{\text{train}}$}] \textbf{(Training Set)}: This is the dataset used to train the initial, polluted model. It is a composite set containing both the clean retain data and the polluted forget data. Formally, $\mathcal{D}_{\text{train}} = \mathcal{D}_{\text{retain}} \cup \mathcal{D}_{\text{forget}}^{\text{polluted}}$.

    \item[\textbf{$\mathcal{D}_{\text{retain}}$}] \textbf{(Retain Set)}: This is the subset of the original training data whose knowledge the model should retain after the unlearning process. This set is entirely clean and unpolluted. A primary goal of any unlearning algorithm is to maintain high performance on $\mathcal{D}_{\text{retain}}$.

    \item[\textbf{$\mathcal{D}_{\text{forget}}$}] \textbf{(Forget Set)}: This is the subset of the original training data whose influence the model must forget. During the initial training, the polluted version of this set, $\mathcal{D}_{\text{forget}}^{\text{polluted}}$, is used. For evaluation after unlearning, the corresponding clean version, $\mathcal{D}_{\text{forget}}^{\text{clean}}$, is typically used to verify that the model can now correctly classify this type of data. It is important to note that for unlearning methods like Gradient Ascent, the polluted version, $\mathcal{D}_{\text{forget}}^{\text{polluted}}$, is used during the unlearning phase to directly counteract the original erroneous learning signals.

    \item[\textbf{$\mathcal{D}_{\text{val}}$}] \textbf{(Validation Set)}: This is a completely clean dataset that is held out from all training and unlearning procedures. It serves as the gold standard for evaluating the final performance of the unlearned model, as it measures the model's ability to generalize to unseen clean data after the unlearning process is complete.
\end{itemize}

\subsection*{Machine Unlearning Methods}

To assess the capacity of each model for repair, we implement and compare four distinct machine unlearning algorithms. Each procedure begins with a model pre-trained on a corrupted dataset and aims to remove the influence of a specified forget set, $\mathcal{D}_{\text{forget}}$, while preserving knowledge from the clean retain set, $\mathcal{D}_{\text{retain}}$.

\textbf{Retrain.}
Considered the gold standard for unlearning, this method involves completely discarding the polluted model and training a new model from a random initialization, using only the clean retain set $\mathcal{D}_{\text{retain}}$. Its objective is to minimize the standard empirical risk on this sanitized data:
\begin{equation}
    \theta_{\text{new}} = \arg\min_{\theta} \sum_{(x,y) \in \mathcal{D}_{\text{retain}}} \mathcal{L}(\theta; x, y)
\end{equation}
While providing a perfect removal of the forget set's influence, this approach is often computationally prohibitive for large-scale models.

\textbf{Finetune.}
This is a computationally efficient approximation of retraining. The procedure starts with the parameters of the already-trained polluted model, $\theta_{\text{old}}$, and continues the training process using only the clean retain set $\mathcal{D}_{\text{retain}}$. The objective is to overwrite the bad knowledge learned from the forget set by leveraging the phenomenon of catastrophic forgetting. The loss function is identical to that of the Retrain method, but the optimization starts from $\theta_{\text{old}}$ instead of a random initialization.

\textbf{Gradient Ascent.}
This method implements a more targeted reverse learning process. It unifies the objectives of retaining and forgetting into a single loss function. The model simultaneously performs standard gradient descent on the retain set while performing gradient ascent on the forget set. This is achieved by minimizing a composite loss where the contribution from the forget set is subtracted:
\begin{equation}
    \mathcal{L}_{\text{GA}}(\theta) = \sum_{(x,y) \in \mathcal{D}_{\text{retain}}} \mathcal{L}(\theta; x, y) - \beta \sum_{(x,y) \in \mathcal{D}_{\text{forget}}^{\text{polluted}}} \mathcal{L}(\theta; x, y)
\end{equation}
Here, $\beta$ is a hyperparameter that balances the strength of forgetting against retaining. This unified objective often leads to a more stable optimization process than multi-objective techniques.

\textbf{Scrub.}
This technique employs a multi-objective push-pull framework that explicitly separates the goals of retaining and forgetting~\cite{Kurmanji2023scrub}. It uses the original polluted model, $\theta_{\text{old}}$, as a teacher to guide the unlearning of a student model, $\theta$. The total loss is a weighted sum of three distinct components:
\begin{equation}
\begin{split}
    \mathcal{L}_{\text{scrub}}(\theta) = {} & \lambda_{\text{ce}} \mathcal{L}_{\text{CE}}(\theta; \mathcal{D}_{\text{retain}}) \\
    & + \lambda_{\text{kl}} \mathcal{L}_{\text{KL, retain}}(\theta) \\
    & - \lambda_{\text{fo}} \mathcal{L}_{\text{KL, forget}}(\theta)
\end{split}
\end{equation}
The first term is a standard cross-entropy loss on the retain set to maintain accuracy. The second term, $\mathcal{L}_{\text{KL, retain}}$, uses the Kullback-Leibler (KL) divergence to encourage the student's outputs on the retain set to remain close to the teacher's, thereby anchoring useful knowledge. The final term, $\mathcal{L}_{\text{KL, forget}}$, maximizes the KL divergence between the student's and teacher's outputs on the forget set, explicitly pushing the student model away from the teacher's incorrect predictions. The hyperparameters $\lambda_{\text{ce}}$, $\lambda_{\text{kl}}$, and $\lambda_{\text{fo}}$ must be carefully tuned to balance these often-conflicting objectives.

\section*{Acknowledgments}
YQC acknowledges the support by
NSAF No. U2330401. SXZ acknowledges the support
from Innovation Program for Quantum Science and
Technology (2024ZD0301700) and the start-up grant
at IOP-CAS.


\let\oldaddcontentsline\addcontentsline
\renewcommand{\addcontentsline}[3]{}

%

\let\addcontentsline\oldaddcontentsline
\onecolumngrid

\clearpage
\newpage
\widetext

\begin{center}
\textbf{\large Supplemental Materials for ``Superior resilience to poisoning and amenability to unlearning in quantum machine learning''}
\end{center}

\renewcommand{\thefigure}{S\arabic{figure}}
\setcounter{figure}{0}
\renewcommand{\theequation}{S\arabic{equation}}
\setcounter{equation}{0}
\renewcommand{\thesection}{\Roman{section}}
\setcounter{section}{0}
\setcounter{secnumdepth}{4}

\section{Analytical Derivation of Landscape Curvature from Minimal Models}
\label{sec:sm_analytical}

To provide a rigorous theoretical foundation for the empirically observed differences in landscape stability, we analyze minimal yet faithful versions of our classical and quantum models. These simplified models, which still incorporate the core sigmoid activation and binary cross-entropy loss, are analytically tractable and reveal the fundamental mechanism behind the QNN's superior resilience.

\subsection{The Minimal Models}
We consider a single data point $(x, y)$, where the feature $x$ is a scalar and the label $y \in \{0, 1\}$.

\paragraph{Classical Model: Single Neuron with Sigmoid}
This model serves as a one-parameter MLP. The pre-activation output (logit) is given by $z = \theta x$. The predicted probability is then $p = \sigma(z) = (1 + e^{-\theta x})^{-1}$, where $\sigma(\cdot)$ is the sigmoid function. The loss is the binary cross-entropy:
\begin{equation}
    \mathcal{L}_{MLP}(\theta) = -[y \log(p) + (1-y)\log(1-p)].
\end{equation}

\paragraph{Quantum Model: Single-Qubit Classifier with Sigmoid}
This is one of the simplest possible QNNs. The feature $x$ is encoded into a quantum state via a rotation, and a single trainable parameter $\theta$ defines the classification boundary. The pre-activation logit is derived from a quantum measurement, $z = \cos(\pi x + \theta)$. This functional form is a direct consequence of a standard and fundamental single-qubit classifier architecture involving three steps: angle encoding, a parameterized rotation, and a final measurement.

\begin{enumerate}
    \item \textbf{State Encoding:} We encode a normalized scalar feature $x$ into the state of a qubit using angle encoding. Starting from the ground state $|0\rangle$, we apply a rotation gate, for instance $R_Y(\pi x)$. The input state becomes:
    \begin{equation}
        |\psi_{\text{in}}\rangle = R_Y(\pi x) |0\rangle = \begin{pmatrix} \cos(\pi x/2) \\ \sin(\pi x/2) \end{pmatrix}.
    \end{equation}

    \item \textbf{Parameterized Processing:} Next, a learnable, parameterized gate is applied. We use a rotation $R_Y(\theta)$ with a single trainable parameter $\theta$. The output state is $|\psi_{\text{out}}\rangle = R_Y(\theta)|\psi_{\text{in}}\rangle$. This simplifies to:
    \begin{equation}
        |\psi_{\text{out}}\rangle = R_Y(\pi x + \theta)|0\rangle.
    \end{equation}

    \item \textbf{Measurement:} To obtain a classical output logit $z$, we compute the expectation value of a standard observable, the Pauli-Z operator $\sigma_z$. The expectation value is calculated as:
    \begin{equation}
        z = \langle \psi_{\text{out}} | \sigma_z | \psi_{\text{out}} \rangle = \cos^2\left(\frac{\pi x + \theta}{2}\right) - \sin^2\left(\frac{\pi x + \theta}{2}\right)=\cos(\pi x + \theta).
    \end{equation}
    
\end{enumerate}
The probability is subsequently calculated as $p = \sigma(z) = \sigma(\cos(\pi x + \theta))$. The loss function is identical to the classical case:
\begin{equation}
    \mathcal{L}_{QNN}(\theta) = -[y \log(p) + (1-y)\log(1-p)].
\end{equation}

\subsection{Analytical Derivation of the Hessian}
The Hessian, $\mathbf{H} = \frac{d^2\mathcal{L}}{d\theta^2}$, represents the local curvature of the loss landscape. We use the chain rule for its derivation. A key property of the cross-entropy loss is that its gradient with respect to the parameter $\theta$ simplifies to:
\begin{equation}
    \frac{d\mathcal{L}}{d\theta} = \frac{d\mathcal{L}}{dp} \frac{dp}{dz} \frac{dz}{d\theta} = \left(\frac{p-y}{p(1-p)}\right) \cdot [p(1-p)] \cdot \frac{dz}{d\theta} = (p-y) \frac{dz}{d\theta}.
\end{equation}

\subsubsection{Hessian of the Single-Neuron MLP}
For the MLP, we have $z = \theta x$, which implies $\frac{dz}{d\theta} = x$. The gradient is therefore $\frac{d\mathcal{L}_{MLP}}{d\theta} = (\sigma(\theta x) - y)x$. Differentiating this expression again with respect to $\theta$ yields the Hessian:
\begin{align}
    \mathbf{H}_{MLP} &= \frac{d}{d\theta}\left[(\sigma(\theta x) - y)x\right] \\
                    &= \left[\frac{d}{d\theta}\sigma(\theta x)\right] \cdot x \\
                    &= [\sigma'(\theta x) \cdot x] \cdot x = x^2 \sigma'(\theta x) \\
                    &= x^2 \sigma(\theta x)(1-\sigma(\theta x)) = x^2 p(1-p).
\end{align}
Since both $x^2$ and the term $p(1-p)$ are always non-negative, the curvature contribution from any single data point to the MLP's loss landscape is strictly non-negative.

At first glance, the expression $\mathbf{H}_{MLP} = x^2 p(1-p)$ appears to be independent of the label $y$. However, this dependence is implicit and strong. The Hessian is evaluated at the optimal parameter $\theta^*$ found during training, and $\theta^*$ is chosen specifically to minimize the loss for a given label $y$. For a clean dataset, the model finds a $\theta^*$ that pushes the prediction $p = \sigma(\theta^*x)$ towards saturation ($0$ or $1$), a region where the term $p(1-p)$ is very small, resulting in a relatively flat minimum. When the training set contains conflicting labels for the same feature $x$, the model is forced to find a compromise parameter $\theta^*$ that results in an uncertain prediction ($p \approx 0.5$). It is precisely in this region of high uncertainty that the term $p(1-p)$ is maximized, leading to a sharp increase in the Hessian's trace. Thus, label noise forces the MLP into a region of high curvature.

\subsubsection{Hessian of the Single-Qubit QNN}
For the QNN, we have $z = \cos(\pi x + \theta)$, which implies $\frac{dz}{d\theta} = -\sin(\pi x + \theta)$. The gradient is $\frac{d\mathcal{L}_{QNN}}{d\theta} = (\sigma(\cos(\pi x + \theta)) - y)(-\sin(\pi x + \theta))$. To compute the Hessian, we differentiate this product using the product rule:
\begin{align}
    \mathbf{H}_{QNN} &= \frac{d}{d\theta}\left[(p-y)(-\sin(\pi x + \theta))\right] \\
                    &= \left[\frac{d(p-y)}{d\theta}\right](-\sin(\pi x + \theta)) + (p-y)\left[\frac{d(-\sin(\pi x + \theta))}{d\theta}\right] \\
                    &= \left[\frac{dp}{d\theta}\right](-\sin(\pi x + \theta)) - (p-y)\cos(\pi x + \theta).
\end{align}
We substitute the chain rule expansion for $\frac{dp}{d\theta} = \frac{dp}{dz}\frac{dz}{d\theta} = p(1-p)(-\sin(\pi x + \theta))$:
\begin{align}
    \mathbf{H}_{QNN} &= \left[p(1-p)(-\sin(\pi x + \theta))\right](-\sin(\pi x + \theta)) - (p-y)\cos(\pi x + \theta) \\
                    &= \underbrace{p(1-p)\sin^2(\pi x + \theta)}_{\text{Term A: Information-Geometric Term}} - \underbrace{(p-y)\cos(\pi x + \theta)}_{\text{Term B: Error-Phase Interaction Term}}.
\end{align}
The QNN's Hessian is composed of two distinct parts. Term A, similar to the MLP's Hessian, is always non-negative. However, Term B's sign depends on both the prediction error $(p-y)$ and the quantum phase $\cos(\pi x + \theta)$. For a mislabeled point that the model correctly identifies as an outlier (e.g., $p \approx 1$ but $y=0$), Term B can become strongly negative.

\subsection{Conclusion from the Analytical Model}
This derivation reveals a fundamental asymmetry. The MLP's landscape is constructed from purely positive curvatures, forcing it to deform to fit noise. The QNN's landscape, due to the oscillatory nature of quantum measurements, contains a noise-adaptive component (Term B) that can introduce negative curvature. This allows the global Hessian to balance or cancel the influence of outliers, leading to a more stable landscape geometry. This analytical insight provides a rigorous theoretical mechanism that underpins the superior resilience and unlearning plasticity observed in our numerical experiments.

\section{Robustness of Learning Patterns Across Model Sizes}
\label{sec:sm_capacity}

A potential concern regarding our main findings is whether the observed differences in resilience are merely an artifact of the specific model sizes chosen for comparison, rather than an intrinsic difference between the classical and quantum learning paradigms. To address this, we conducted a systematic study where we varied the capacity (or expressiveness) of both the MLP and QNN models and re-evaluated their performance under the label flipping corruption protocol on quantum XXZ dataset.

\begin{figure}[h!]
    \centering
    \includegraphics[width=0.9\columnwidth]{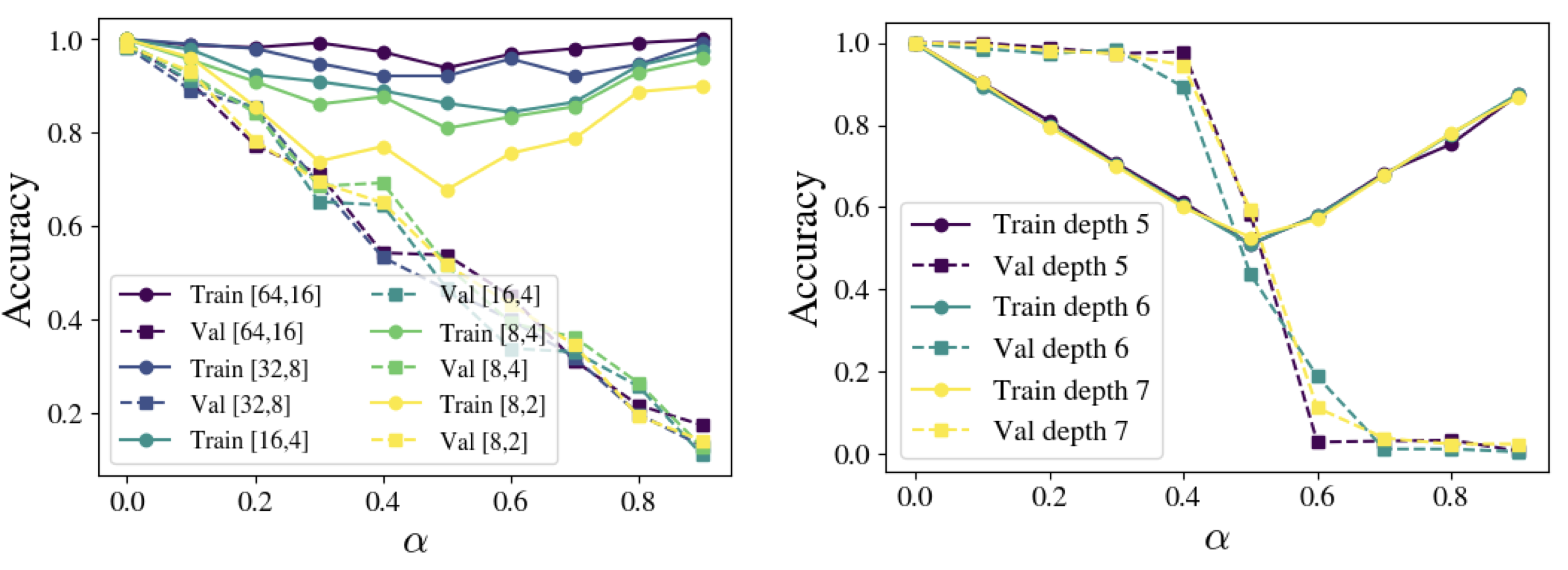}
    \caption{\textbf{Resilience patterns are consistent across different model capacities.} Training (solid lines) and validation (dashed lines) accuracy as a function of the label flip noise ratio $\alpha$ for models of varying sizes.
    Left: Performance of MLPs with different hidden layer sizes (indicated in the legend). All models, regardless of capacity, exhibit the characteristic continuous compromise, where validation accuracy steadily degrades with increasing data noise.
    Right: Performance of QNNs with different circuit depths. All models, regardless of depth, display the distinct robust plateau followed by a critical transition, demonstrating that this behavior is an intrinsic feature of the quantum learning model.}
    \label{fig:sm_capacity}
\end{figure}

For the classical MLP, we varied its capacity by changing the number of neurons in its two hidden layers, ranging from a larger model with [64, 16] neurons to a much smaller one with [8, 2] neurons. For the quantum QNN, we varied its capacity by adjusting the depth of the parameterized quantum circuit, testing depths of 5, 6, and 7. The results of these experiments are presented in Fig.~\ref{fig:sm_capacity}.

The results unequivocally demonstrate that the learning patterns are an intrinsic property of the model class, not its size. As shown in Fig.~\ref{fig:sm_capacity}(left), every MLP, from the largest to the smallest, exhibits the same qualitative behavior: a continuous compromise where validation accuracy begins to degrade immediately with the introduction of noise. While larger MLPs can better memorize the noisy training labels (as seen by their higher training accuracy), their generalization performance is similar to smaller MLPs.

Conversely, Fig.~\ref{fig:sm_capacity}(right) shows that all QNNs, regardless of their circuit depth, display the characteristic pattern of resilience we identified in the main text. Each QNN maintains a robust performance plateau for a significant range of noise ratios before undergoing a sharp critical transition. This demonstrates that the QNN's ability to disregard noise and preserve a generalizable solution is not an accident of a specific parameter count but is fundamental to its learning nature.

Therefore, the stark qualitative difference between the behavior shown in Fig.~\ref{fig:sm_capacity} confirms that our main conclusion is robust. The contrast between brittle memorization and resilient generalization is an intrinsic feature distinguishing the classical and quantum learning paradigms investigated, not a mere consequence of model expressiveness or parameter count.

\section{Unlearning Efficacy on the Forget Set}
\label{sec:sm_forget_set}

To complement the validation accuracy results presented in the main text, we here provide a direct measure of unlearning efficacy on the forget set, $\mathcal{D}_{\text{forget}}$. For this analysis, we define a metric we call forgetting accuracy, which is the accuracy of the model when evaluated on the \textbf{polluted version} of the forget set, i.e., the set with the intentionally flipped labels that the original model was trained on.

A successful unlearning process should cause the model to forget the erroneous associations from the poisoned data. Therefore, a lower Forgetting Accuracy indicates a more effective unlearning procedure. A value of 1.0 means the model still perfectly remembers the incorrect, poisoned labels, while a value of 0.0 means the model has completely forgotten them and now classifies the samples based on their true features, which contradict the poisoned labels. 

\begin{figure}[h!]
    \centering
    \includegraphics[width=0.9\columnwidth]{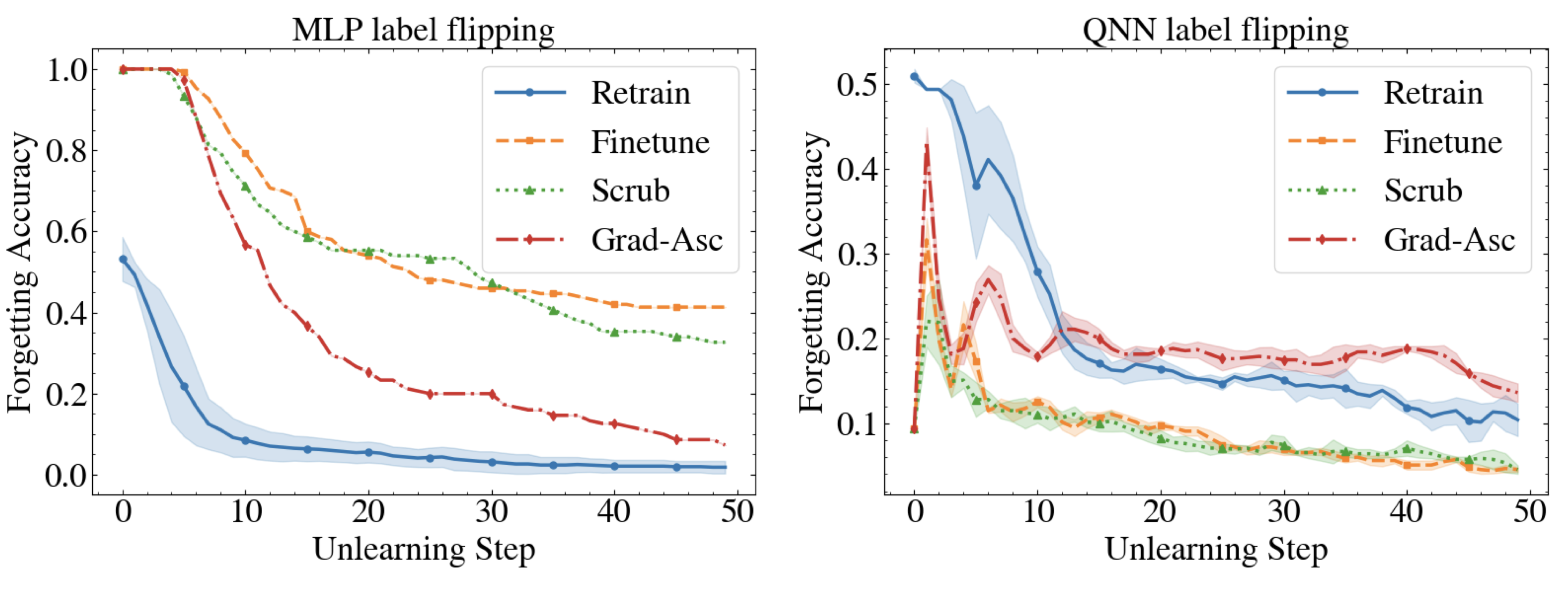}
    \caption{\textbf{Forgetting Accuracy on the MNIST dataset.} The plot shows the evolution of forgetting accuracy on the polluted forget set versus the unlearning step. A rapid decrease to zero indicates effective unlearning. }
    \label{fig:forget_mnist}
\end{figure}

\begin{figure}[h!]
    \centering
    \includegraphics[width=0.9\columnwidth]{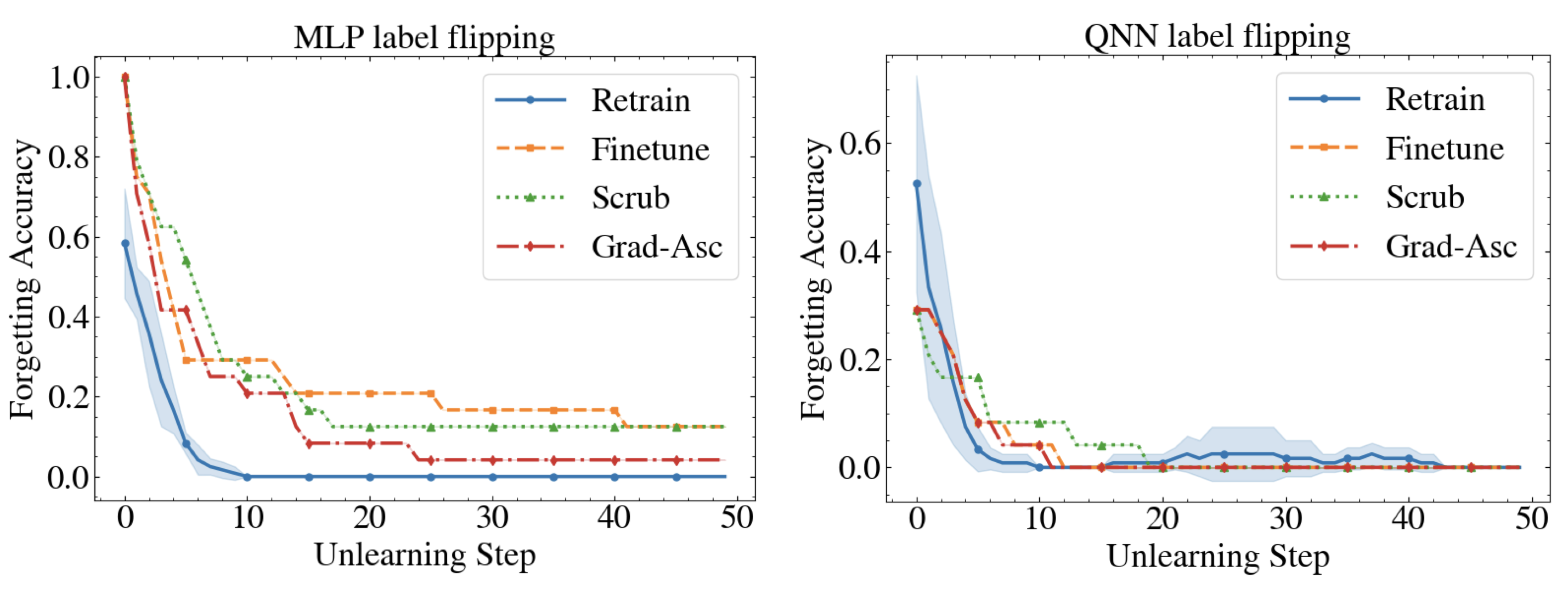}
    \caption{\textbf{Forgetting Accuracy on the XXZ dataset.} The trends are consistent with the MNIST results. The MLP exhibits slow forgetting, indicative of stubborn memories. The QNN demonstrates rapid forgetting across all methods, indicative of high model plasticity.}
    \label{fig:forget_xxz}
\end{figure}

The results, presented in Fig.~\ref{fig:forget_mnist} for the MNIST dataset and Fig.~\ref{fig:forget_xxz} for the XXZ dataset, provide direct evidence for the conclusions drawn in the main text.

For the classical MLP, we observe that while the Retrain baseline quickly achieves a forgetting accuracy near zero, the approximate methods lag significantly behind. Their forgetting accuracy decreases slowly and remains substantially above zero even after 50 unlearning steps. This large gap between the approximate methods and the baseline is a clear signature of the stubborn memories formed by the MLP; the incorrect knowledge is deeply entrenched and difficult to erase efficiently.

In contrast, the QNN demonstrates a much higher degree of model plasticity. All unlearning methods, including the approximate ones, show a rapid and dramatic decrease in forgetting accuracy. This indicates that the QNN's memory of the corrupted data is far less rigid and can be efficiently and effectively erased.
These findings directly support our central thesis: the QNN's inherent resilience and the stable geometry of its loss landscape make it not only resistant to initial corruption but also highly amenable to subsequent correction.

\section{Hyperparameters in architectures and setups}
We evaluate and compare an MLP and a QNN on machine learning and unlearning tasks. The model architectures and hyperparameters for both the initial training and subsequent unlearning phases are detailed below.

\subsection{Model Architectures}

\paragraph{Classical Model (MLP):}
The MLP is a fully connected neural network utilizing ReLU activation functions for hidden layers and a final Sigmoid activation for binary classification.
\begin{itemize}
    \item \textbf{Hidden Units Number:} The architecture of the MLP varied across experiments.
    \begin{itemize}
        \item For the  \texttt{XXZ} dataset, the MLP consists of two hidden layers with 64 and 16 units, respectively.
        \item For the \texttt{MNIST} dataset, a smaller MLP with 16 and 4 hidden units was used.
    \end{itemize}

\end{itemize}

\paragraph{Quantum Model (QNN):}
The QNN is built on top of variational quantum circuits.
\begin{itemize}
    \item \textbf{Circuit Depth:} The depth of the quantum circuit was also varied.
    \begin{itemize}
        \item For the \texttt{XXZ} experiments, a circuit depth of 4 and 12 qubits was used.
        \item For the \texttt{MNIST} experiments, a circuit depth of 6 and 10 qubits was used.
    \end{itemize}
\end{itemize}

\subsection{Initial Training Process}

The models were first trained on a dataset, which includes poisoned or mislabeled samples. All experiments were averaged over 5 independent runs for statistical robustness. The Adam optimizer was used for training all models. The specific hyperparameters for the two main experimental sets are detailed in Table~\ref{tab:training_params}.

\begin{table}[h!]
\centering
\large
\caption{Hyperparameters for the initial training process.}
\label{tab:training_params}
\begin{tabular}{|l|c|c|}
\hline
\multicolumn{3}{|c|}{\textbf{\texttt{XXZ} Dataset Experiments}} \\ \hline
\textbf{Parameter} & \textbf{MLP} & \textbf{QNN} \\ \hline
Epochs & 400 & 200 \\
Batch Size & 32 & 32 \\
Learning Rate & 0.01 & 0.03 \\ \hline
\multicolumn{3}{|c|}{\textbf{\texttt{MNIST} Dataset Experiments}} \\ \hline
\textbf{Parameter} & \textbf{MLP} & \textbf{QNN} \\ \hline
Epochs & 100 & 100 \\
Batch Size & 128 & 256 \\
Learning Rate & 0.01 & 0.005 \\ \hline
\end{tabular}
\end{table}

\subsection{Unlearning Process and Hyperparameters}

After initial training, various unlearning methods (\texttt{retrain}, \texttt{finetune}, \texttt{scrub}, \texttt{gradient ascent}) were applied.

\begin{itemize}
    \item \textbf{Definition of an Unlearning Step:} An ``unlearning step'' is defined as a single, full epoch. The batch size for each unlearning step is set to the entire size of the retain set. This means the model processes all retain data in one pass for each step.
    \item \textbf{Unlearning Steps:} The unlearning process was run for 50 steps (\texttt{Epochs=50}).
    \item \textbf{Unlearning Optimizer/Hyperparameters:} The unlearning process also utilizes the Adam optimizer. The learning rate  was set to 0.01 for the \texttt{XXZ} experiments and 0.001 for the \texttt{MNIST} experiments.
    \item \textbf{Unlearning Loss Weights:} The unlearning loss function is a weighted sum of multiple components, including a KL-divergence term ($\lambda_{kl}$), a cross-entropy term ($\lambda_{ce}$), and a forgetting term ($\lambda_{fo}$ or $\beta$). After preliminary experiments, we selected a set of weights that demonstrated strong performance and the specific weight configurations for each method are detailed below:

    \begin{itemize}
        \item Retrain: This method involves retraining the model from scratch on the retain set only. The loss function is the standard cross-entropy on the retain set, effectively setting the unlearning-specific weights as: $\lambda_{ce} = 1.0$.

        \item Finetune: This method finetunes the pre-trained model on the retain set. Similar to Retrain, the loss function is the standard cross-entropy, with the weights set to:  $\lambda_{ce} = 1.0$.

        \item Scrub: This method is a specific unlearning algorithm tested with the following weights: $\lambda_{kl} = 0.0$, $\lambda_{ce} = 1.0$, $\lambda_{fo} = 0.2$.

        \item Grad-Asc: This gradient-based method was tested with the same weight configuration as the Scrub method: $\lambda_{ce} = 1.0$, $\beta = 0.2$.
    \end{itemize}
\end{itemize}

\subsection{Hessian Matrix Evaluation}

We use \texttt{MNIST} classification task to evaluate Hessian matrix for trained models. For methods requiring the computation of the Hessian matrix, a subset of the training data is used to make the calculation computationally tractable. The size of this subset is explicitly set to 100 samples. These samples are typically chosen randomly from the training set for each evaluation.

\begin{itemize}
    \item \textbf{Training Data:} For each of the two classes of digits, we randomly select 250 samples.
    \item \textbf{Preprocessing:} The original $28 \times 28$ images  are flattened into vectors. To accommodate the input requirements of the quantum model, these vectors are zero-padded to a dimension of 1024  and then L2-normalized.
    \item \textbf{Noise Injection:} To simulate a data poisoning scenario, label noise is introduced into the training set. The experiments are conducted across multiple noise levels, with noise ratios $\alpha$ of 20\%, 30\%, and 40\%.
\end{itemize}


\end{document}